\documentclass{iaus}
\usepackage{graphicx}

\title[Progenitor's signatures in Type Ia SNRs] 
{Progenitor's signatures\\ in Type Ia supernova remnants}

\author[A. Chiotellis, D. Kosenko, K. M. Schure   \& J. Vink]   
{A. Chiotellis$^1$,
\ D. Kosenko$^{1}$, 
\ K. M. Schure$^{1,2}$
\and J. Vink$^1$}

\affiliation{$^1$Sterrenkundig Instituut  Utrecht, \\ Postbus 80000,
NL-3508TA, Utrecht, the Netherlands \\ email: {a.chiotellis@astro-uu.nl} \\[\affilskip]
$^2$Department of Physics, University of Oxford, \\
             Clarendon Laboratory, Parks Road, Oxford OX1 3PU, United Kingdom }

\pubyear{2011}
\volume{281}  
\pagerange{1--4}
\jname{Binary Paths to Type Ia Supernovae Explosions}
\editors{R. Di Stefano \& M. Orio}
\begin{document}

\maketitle

\begin{abstract}

The remnants of Type Ia supernovae can provide important clues about their progenitor-histories. 
 We discuss two well-observed supernova remnants (SNRs) that are believed to result from a Type Ia SN and use various tools to shed light on the possible progenitor history. We find that Kepler's SNR  is consistent with a symbiotic binary progenitor consisted of a white dwarf and an AGB star. Our hydrosimulations can reproduce the observed kinematic and morphological properties. For Tycho's remnant we use the characteristics of the X-ray spectrum and the kinematics to show that the ejecta has likely interacted with dense circumstellar gas.

\end{abstract}

\firstsection 
\section{Introduction}

 Type Ia supernovae (SNe Ia) have been  the key objects for the discovery that the universe is accelerating. In addition, they are one of the main sources of the chemical enrichment in galaxies with iron peak elements. 
Given their importance, it is disconcerting that their nature is still poorly understood.  SNe Ia are believed to result from the thermonuclear explosion of a CO white dwarf (CO WD) in binary systems which approach the Chandrasekhar mass through mass accretion from the companion star. However, the nature of the donor star, the  binary evolution path that leads to SNe Ia and the explosion mechanism are still unclear. 
Different evolutionary paths of Type Ia progenitors lead to different modifications of the ambient medium, either through mass outflows, or through   ionizing radiation that can accompany accretion.
The subsequent interaction of the supernova ejecta with the modified (or not) circumstellar medium (CSM)  leads to different properties of the SNRs (morphology, dynamics, spectra etc.). Thus, the local population of Type Ia SNRs can provide us with valuable information about Type Ia progenitors.
  
 Here we  model  two historical SNe: SN1604 (Sect.2) and SN1572 (Sect. 3). In both cases we study the impact of the interaction between the SN ejecta and dense circumstellar structures on the observational properties of these SNRs.

\section{The case of Kepler's SNR (SN 1604)}

Kepler's SN occured in 1604  high above the Galactic plane (G4.5+6.8). Its radius is 2.6$d_5$~pc, with $d_5$ the distance in units of 5 kpc. 
This SNR has been a puzzling object due to its increased emissivity  
in the northern region, which shows a substantial overabundance of nitrogen, ${\rm N/N_{\odot} > 2}$, but otherwise solar metallicity (\cite[Blair et al. 1991]{Blair91}). 

The presence of the nitrogen-rich  shell appears to affect the dynamics, as in the north the expansion parameter ($m = (dR/dt)/(R/t)=0.35$) is lower than the rest of the SNR, m= 0.6 (\cite[Vink 2008]{Vink08}) and lower than expected for young SNRs ($m>0.4$).  All of the aforementioned properties indicate  the existence of a massive shell in the northern region that was formed by mass outflows during the evolution of the progenitor system.

We have shown that these characteristics  can be explained if the CSM was  shaped by the stellar wind of an asymptotic giant branch (AGB) donor star (\cite[Chiotellis et al. 2011]{Chiotellis11}). 
AGB stars with initial masses $ > 4 {\rm M_{ \odot}}$ are able to enrich their surfaces with nitrogen. Based on the AGB models of \cite{Karakas07}, the chemical composition of the circumstellar shell at Kepler's SNR can be best explained if the AGB donor star had an initial mass of $ 4 -5 {\rm M_{ \odot}}$ and solar metallicity.   This suggests that the progenitor system was  a wide symbiotic binary, where part of the  slow wind of the donor has been accreted onto the WD while the rest of the wind formed the observed nitrogen-rich shell.  Finally, we retain the idea, first suggested by \cite{Bandiera87}, that the asymmetry of the northern shell of Kepler's SNR  can be explained by its observed supersonic motion of $250 ~{\rm km~s^{-1}}$ away from the Galactic plane (\cite[Bandiera \& van den Bergh 1991; Sollerman et al. 2003]{Bandiera91,Sollerman03}). The interaction of the stellar wind with the ram pressure of the ISM leads to the formation of a bow-shaped shell. Nowadays, the SNR's  blast wave  interacts only with the nearest region of this bow shell.  

\begin{figure}[hbt]
\begin{center}
 \includegraphics[trim=55 25 320 45,clip=true,width=30mm,angle=0]{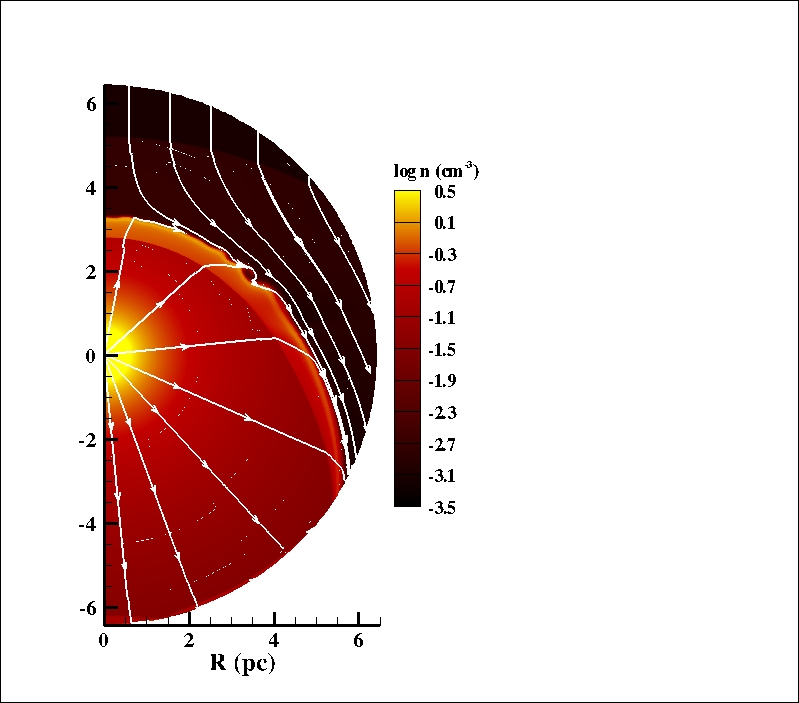} 
 \includegraphics[trim=55 25 320 45,clip=true,width=30mm,angle=0]{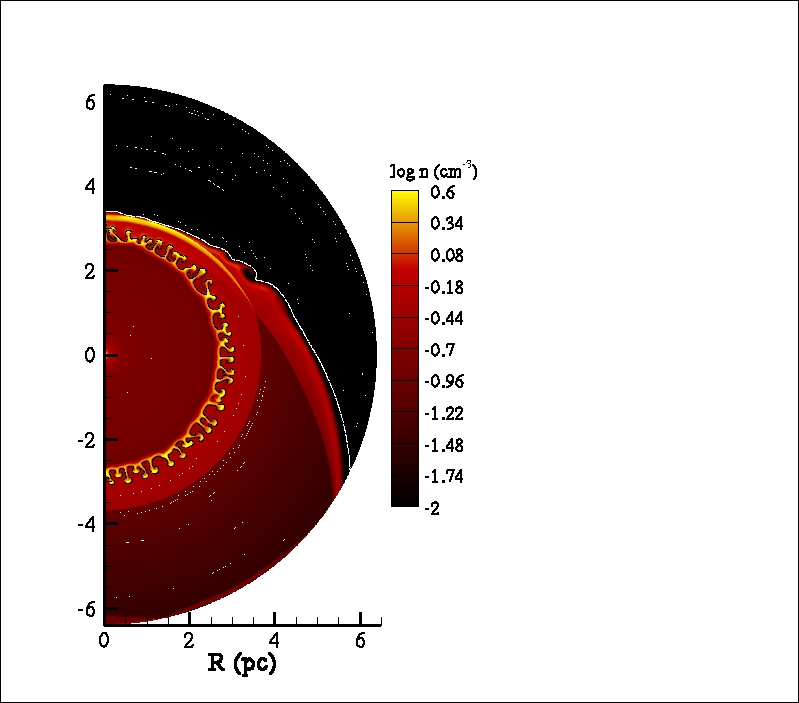} 
 \includegraphics[trim=55 25 320 45,clip=true,width=30mm,angle=0]{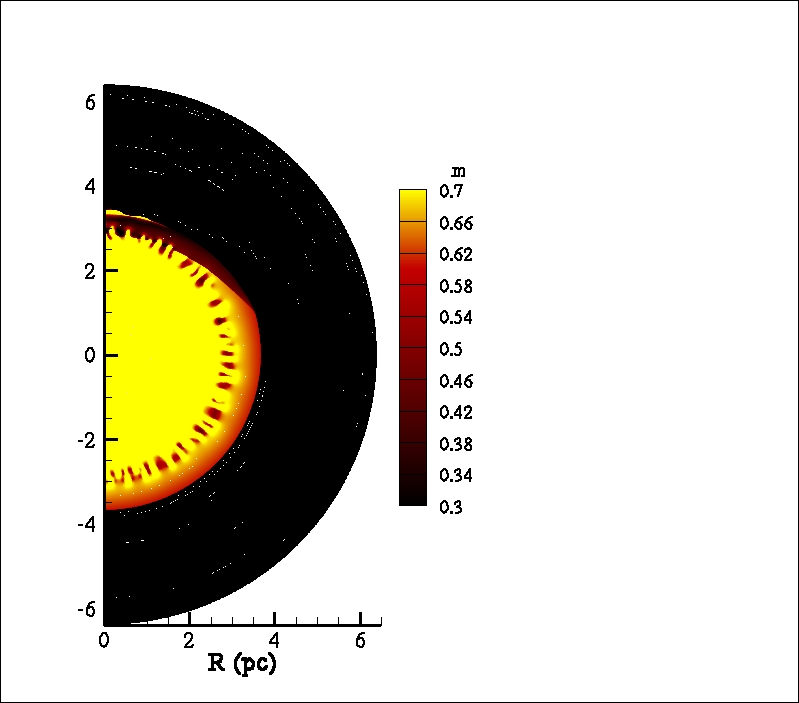} 
  \caption{Left: The formation of the bow-shaped shell. The wind parameters that have been used are: mass loss rate $ \dot{M}= 10^{-5} ~{\rm M_{\odot} yr^{-1}}$, wind velocity: $u_w = 10 ~{\rm km~ s^{-1}}$, ISM density $n_{ism} = 7 \times 10^{-4} {\rm cm^{-3}}$, systemic velocity  $u_* = 250~ {\rm km~ s^{-1}}$ while the timescale of the bubble evolution is 0.38 Myr. The density and the expansion parameter of the SNR at the current age of Kepler's SNR are depicted in the middle and right panel respectively. The energy of the SN is $10^{51}$ erg while the mass of the ejecta $1.4 {\rm M_{\odot}}$ (\cite[Chiotellis et al. 2011]{Chiotellis11}).}
   \label{fig1}
\end{center}
\end{figure}

In order to test  this scenario we have performed 2D hydrosimulations employing the  AMRVAC code (\cite[Keppens et al. 2003]{Keppens03}). 
First we simulate the formation of the CSM by imposing an inflow with the properties of a spherical, cold and slow stellar wind. At the same time the ISM with constant density ($\rho_{ism}$) enters from one side with momentum $m= \rho_{ism}  u_*$, to represent the systemic motion of the progenitor system with velocity $u_*$ (see Fig. \ref{fig1})  and forms the bow-shaped shell. In the second stage, we introduce the supernova ejecta and let the SNR evolve.  Fig. \ref{fig1} shows the result of the simulation  at the current age of Kepler's SNR.  The model reproduces the observed characteristics of Kepler's SNR: The remnant interacts with the nearest region of the bow shell only, explaining the observed asymmetry,  and has an expansion rate inside the shell of $m = 0.3 - 0.35$ versus $m=0.6$ in the rest of the remnant, in agreement with the observations.

\section{The case of Tycho's SNR (SN 1572)}
 The case for a CSM shaped by the progenitor for Kepler is quite obvious. But what about other Type Ia
SNRs? For another important historical SNR, SN 1572 /Tycho's SNR the case for
a CSM shell is less conspicuous. However,
there is a clear discrepancy between the ISM densities as measured
by modeling the X-ray emission from a delayed-detonation explosion (\cite[Badenes et al. 2006]{Badenes06}),  $2 \times 10^{-24} {\rm g~cm^{-3}} $,
and density estimates based on the kinematics of the SNR. The latter indicating
a density that is five times lower  (\cite[Katsuda et al. 2010]{Katsuda10}). For this reason \cite{Katsuda10} have suggested that a more complex circumstellar stucture may resolve this discrepancy.  

 In order to study the effects of a dense shell we have simulated the X-ray emission from two SNRs;  one evolving into a homogeneous ISM, and the other shortly interacting with a CSM that was formed by a stellar wind, before further propagating into a homogeneous ISM. 

\begin{figure} [htbp]
\begin{center}$
\begin{array}{ccc}
\includegraphics[trim=0 0 0 0,clip=true,height=31.5mm,angle=0]{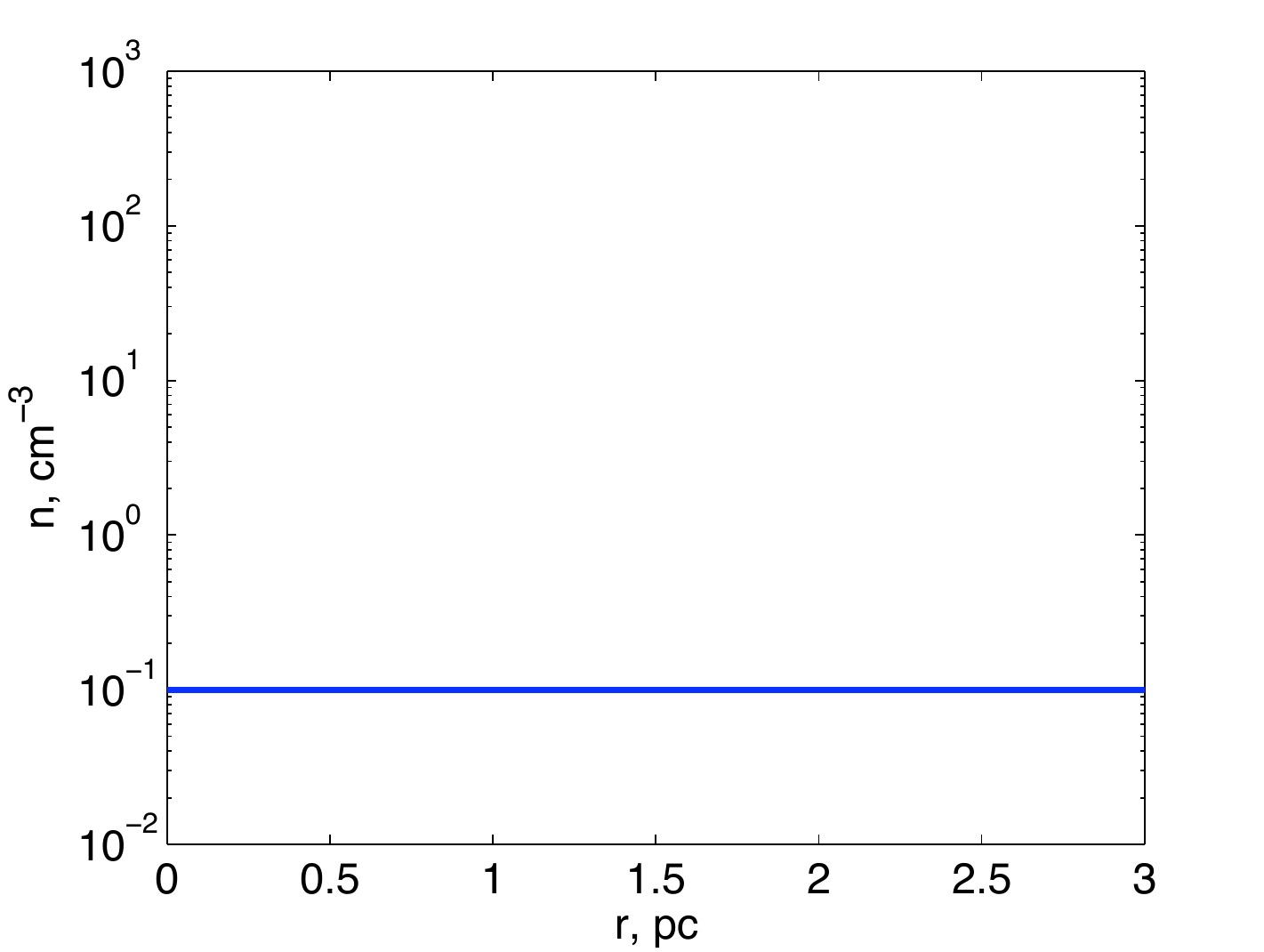}&
\includegraphics[trim=0 0 0 0,clip=true, height =30mm,angle=0]{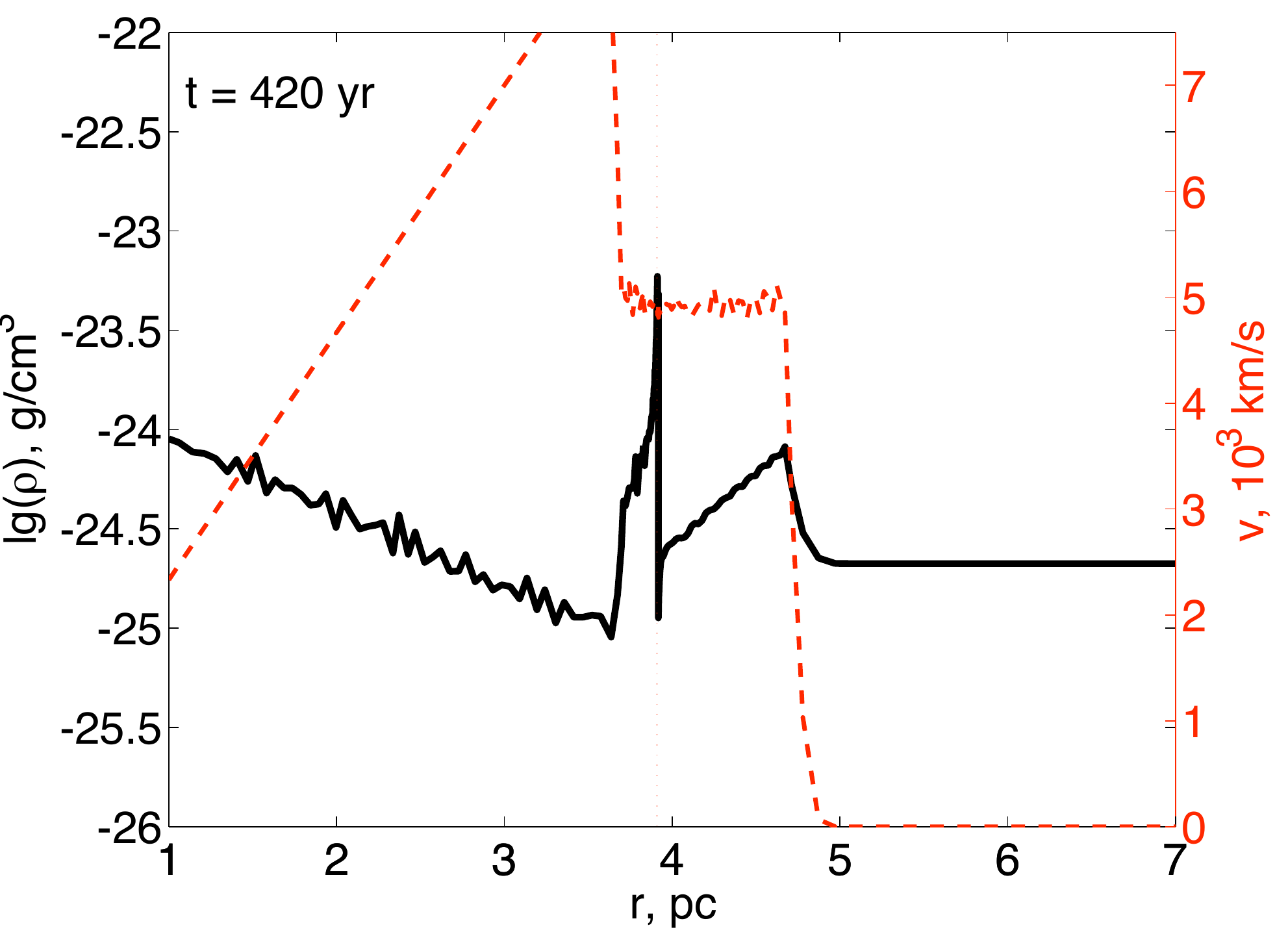}&
\includegraphics[trim=0 0 0 0,clip=true, height =30mm,angle=0]{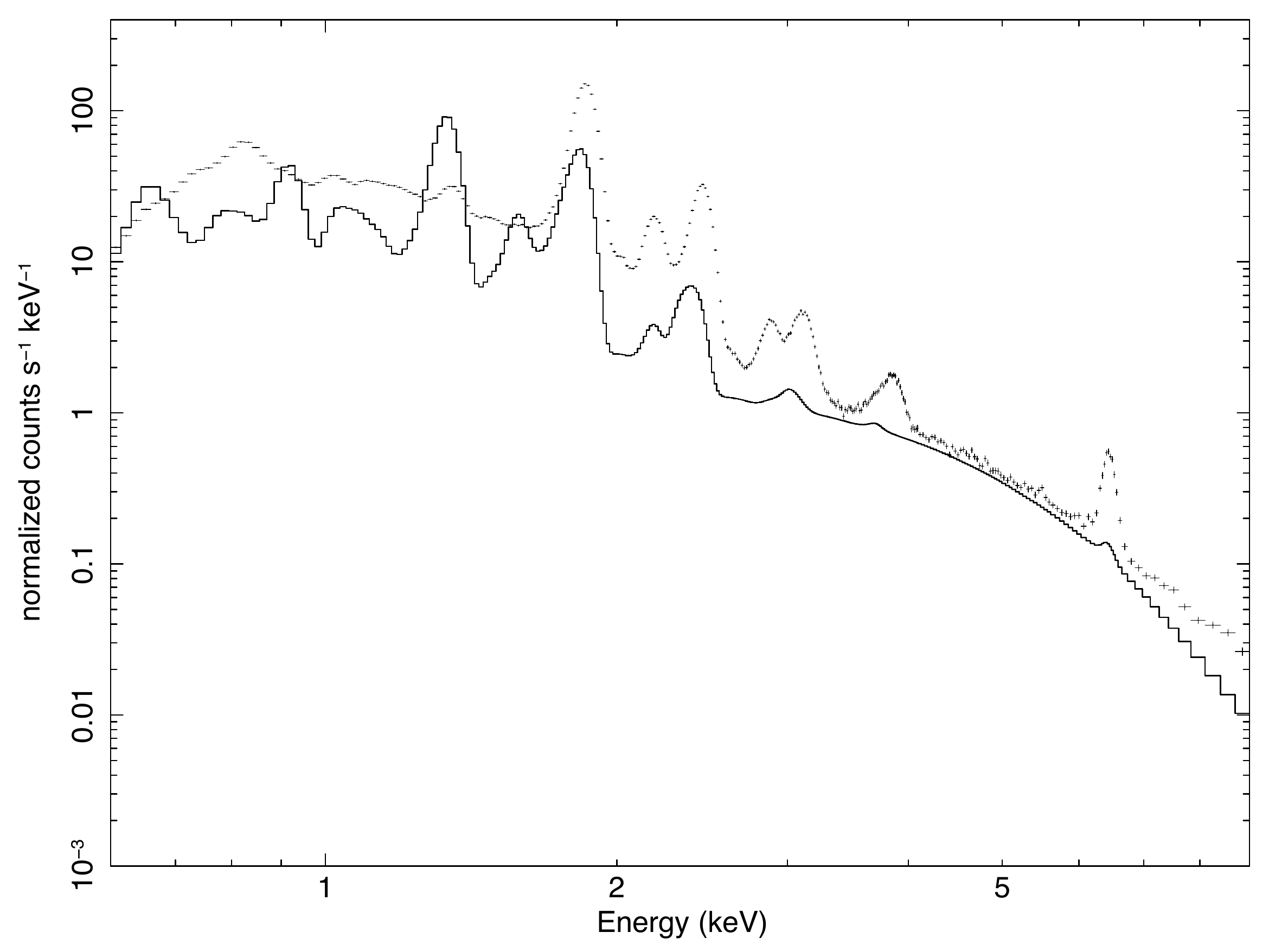}
\\[-1mm]
\includegraphics[trim=0 0 0 0,clip=true, height =31.5mm,angle=0]{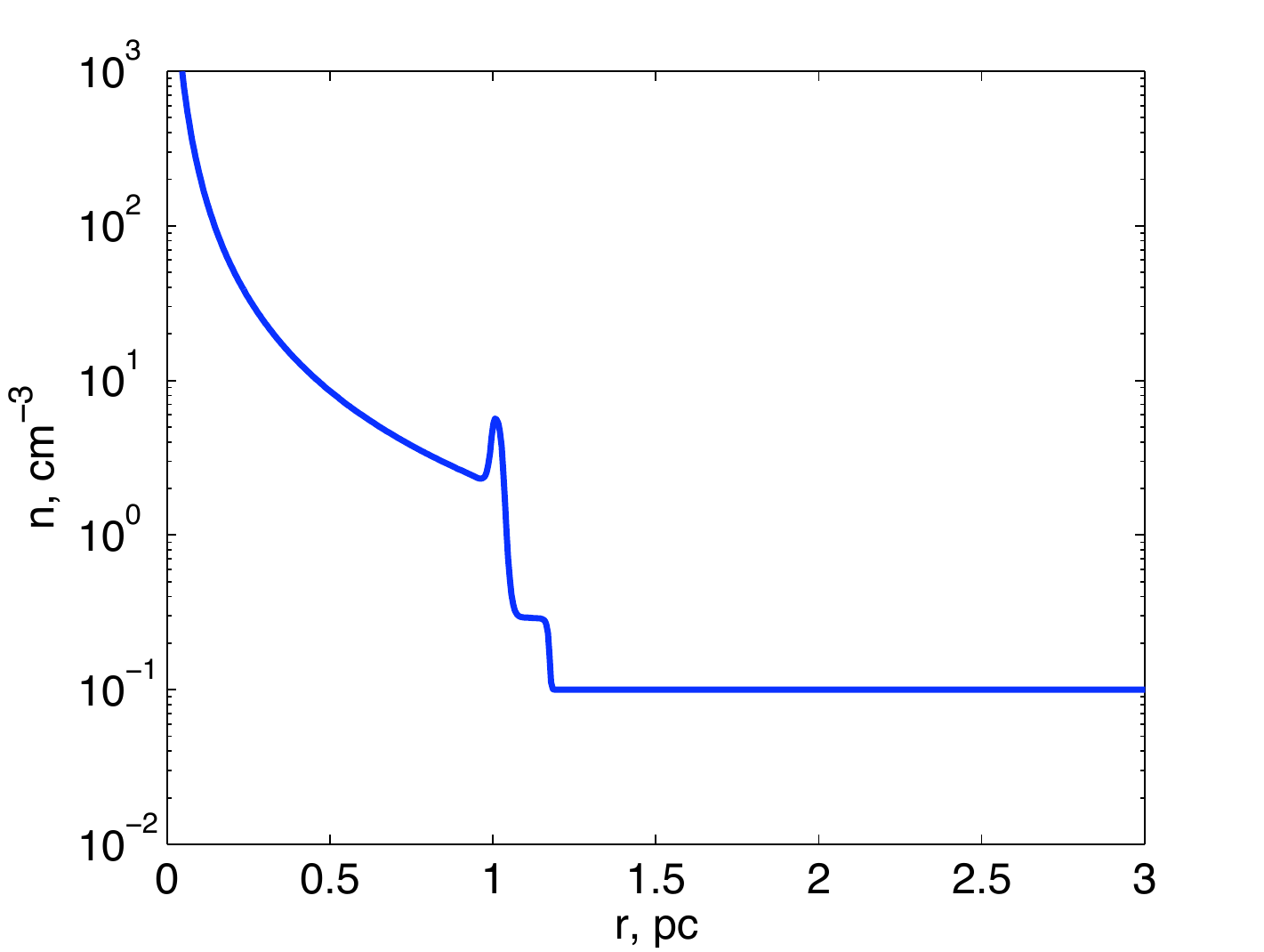}& 
\includegraphics[trim=0 0 0 0,clip=true, height =30mm,angle=0]{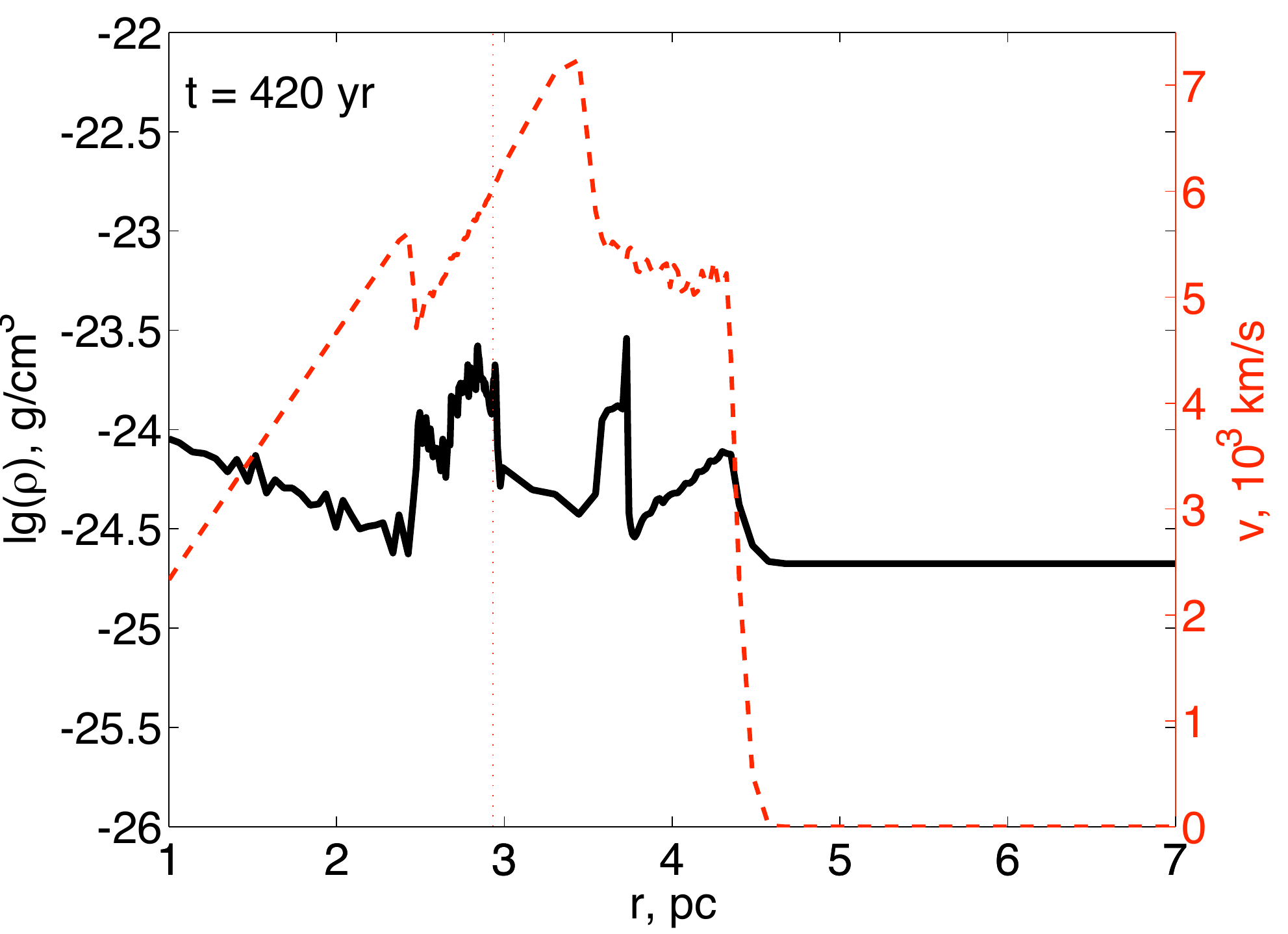}&
\includegraphics[trim=0 0 0 0,clip=true, height =30mm,angle=0]{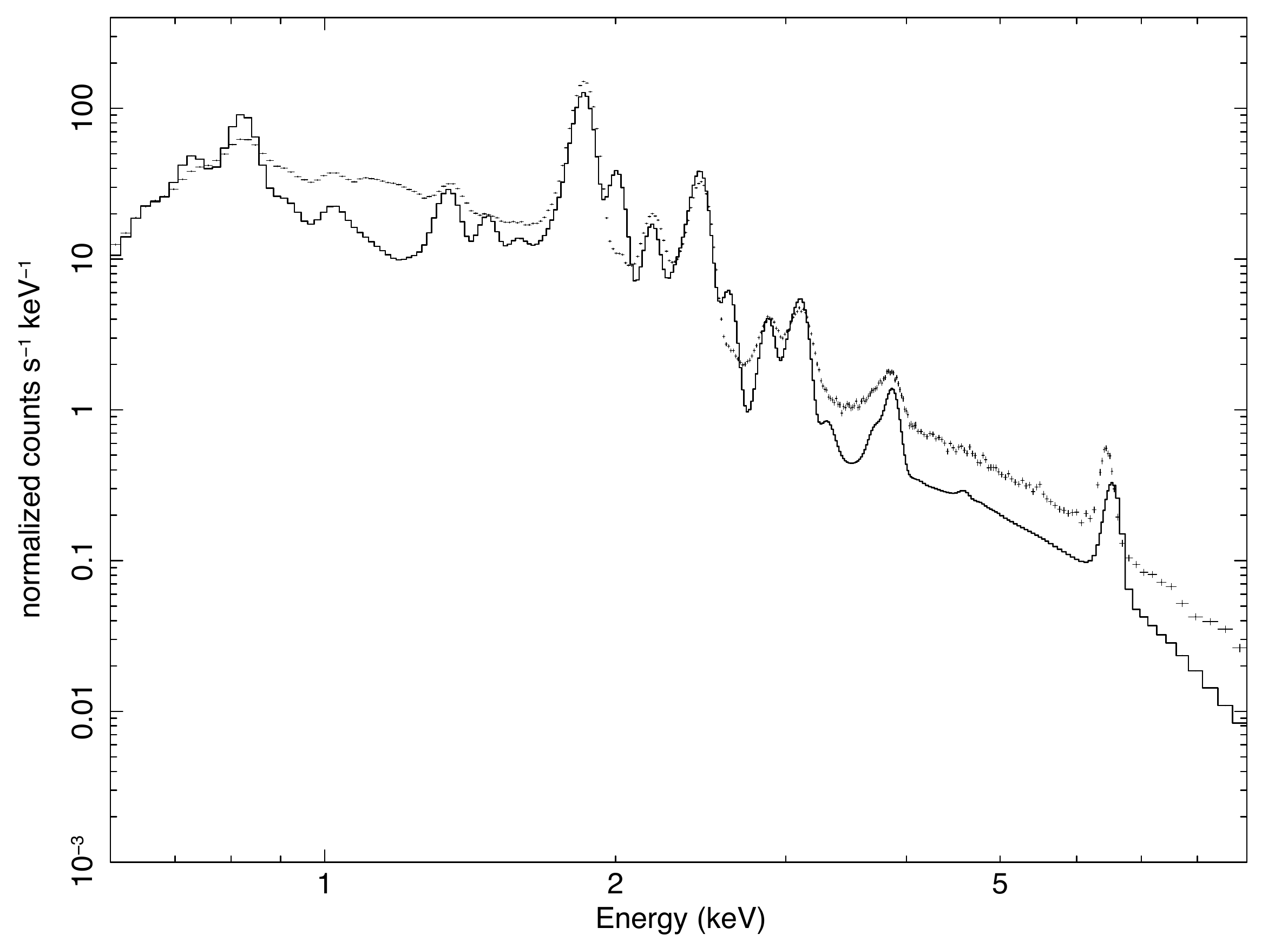} 
\end{array}$
\end{center}
\caption{Left: the number density profiles of the ambient medium before the explosion. Middle: the properties of the SNR at the current age of Tycho's SNR (black solid line is the mass density, red dashed line is the velocity profile, vertical dotted line indicates the location of contact discontinuity). Right: the X-ray spectrum for each model (solid lines) in comparison with  the observed XMM-Newton  spectra of Tycho's SNR (crosses).  Note that X-ray synchrotron emission contributions (continuum) has not been included here. Top row: the case of the SNR evolution in a homogeneous ISM. Bottom row: the case of 
a SNR interacting  with a wind bubble. The wind parameters are: mass loss rate $ \dot{M}= 10^{-5} \;{\rm M_{\odot} yr^{-1}}$, wind velocity $u_w = 10 \;{\rm km\,s^{-1}}$ and the wind outflow phase is 0.1 Myr. In both cases the ISM density is  $\rho = 2\times10^{-25}\;{\rm g\,cm^{-3}}$ (Kosenko et al. in preparation). }
 \label{SNR_spectra}
\end{figure}

We performed our simulations in three steps. Initially, using the AMRVAC code, we form the ambient medium for the wind profile (first column of Fig.~\ref{SNR_spectra}). Subsequently, we let the SNR evolve in either a homogeneous ISM, or in the CSM+ISM, using the hydrodynamical code SUPREMA (\cite{sorokina:04}, \cite{kosenko:11}) with the W7 deflagration explosion model (\cite{nomoto:84}). Finally, we calculate the X-ray emission from the simulated SNRs employing the SPEX software package (\cite{kaastra:96}).

 We produced a number of simulations with various wind parameters. Fig.~\ref{SNR_spectra} shows a typical example that fits the characteristics of Tycho's SNR well. In the case of the SNR expanding in the CSM the density and velocity structures  are more complicated in comparison with the classical ISM case. The swept up mass of the CSM is about $2.0\; {\rm{M_\odot}}$, in the ISM case the swept up mass is $\sim1.4\; {\rm M_\odot}$. The resulting  thermal X-ray spectra differs drastically due to the different temperature and ionization timescale distributions throughout the shocked supernova ejecta. The fluxes of the emission lines and the locations of  their centroids are defined by these parameters (\cite[for the detailed studies see e.g. Badenes et al. 2006]{Badenes06}). In the specific case of Tycho's SNR, the W7 explosion model in the wind bubble reproduces the observed spectra better (Fig.2, right column), while allowing for densities that give a consistent result for the kinematics. This result indicates that  a non-homogeneous CSM may explain both the kinematics and the spectrum of Tycho SNR.

\section{Discussion }
 While Kepler's SNR shows clear evidence for interaction with CSM, 
for Tycho's SNR the case is more subtle, as it manifests itself in the details of the X-ray spectrum.
The reason may be that in Tycho the SNR blast-wave has already penetrated the shell in the past.

 Based on the kinematics and morpology of the Kepler's SNR and the chemical composition of its northern shell we argue that Kepler's SN had a symbiotic binary progenitor consisting of a CO WD and a $4-5~ {\rm M_{\odot}}$ AGB donor star. 
For the case of Tycho's SNR, the  presence of a dense non-homogeneous ambient medium around its progenitor seems be able to reconcile the differences that result from the studies of the kinematics and the X-ray spectra  of the remnant. The specific structure and origin of this CSM needs further investigation. 

 So here we presented two cases which argue for stellar wind outflows around SNe Ia. There
are several other studies suggesting similar outflows  (e.g. \cite[Sternberg et al. 2011, Borkowski et al. 2006]{Sternberg11,Borkowski:06}). However, direct radio, and X-ray observations of SNe Ia
put rather stringent constraints on outflows from SN Ia progenitors (e.g \cite[Mattila et al. 2005; Panagia et al. 2006; Immler et al. 2006]{Mattila05,Panagia06,Immler06}). Reconciling these discrepancies is an important challenge for future SNe Ia studies.

\end{document}